\documentclass[prd,preprint,superscriptaddress,showpacs,byrevtex]{revtex4}
\usepackage{bm}
\def\fsl#1{\setbox0=\hbox{$#1$}           
   \dimen0=\wd0                                 
   \setbox1=\hbox{/} \dimen1=\wd1               
   \ifdim\dimen0>\dimen1                        
      \rlap{\hbox to \dimen0{\hfil/\hfil}}      
      #1                                        
   \else                                        
      \rlap{\hbox to \dimen1{\hfil$#1$\hfil}}   
      /                                         
   \fi}                                         %
\newcommand{\be}{\begin{equation}}
\newcommand{\ee}{\end{equation}}
\newcommand{\bea}{\begin{eqnarray}}
\newcommand{\eea}{\end{eqnarray}}
\newcommand{\beq}{\begin{equation}}
\newcommand{\eeq}{\end{equation}}
\newcommand{\beqs}{\begin{eqnarray}}
\newcommand{\eeqs}{\end{eqnarray}}

\begin{document}
\title{ Momentum Sum Rule Is Violated in The Operator Product Expansion in QCD At The High Energy Colliders }
\author{Gouranga C Nayak }
\thanks{G. C. Nayak was affiliated with C. N. Yang Institute for Theoretical Physics in 2004-2007.}
\affiliation{ C. N. Yang Institute for Theoretical Physics, Stony Brook University, Stony Brook NY, 11794-3840 USA}
\date{\today}
\begin{abstract}
To prove the momentum sum rule in the operator product expansion (OPE) in QCD at high energy colliders it is assumed that $<P| {\hat T}^{++}(0)|P>=2(P^+)^2$ where $|P>$ is the momentum eigenstate of the hadron $H$ with momentum $P^\mu$ and  ${\hat T}^{++}(0)$ is the $++$ component of the gauge invariant color singlet energy-momentum tensor density operator ${\hat T}^{\mu \nu}(0)$ of all the quarks plus antiquarks plus gluons inside the hadron $H$. However, in this paper, we show that this relation $<P| {\hat T}^{++}(0)|P>=2(P^+)^2$ is correct if ${\hat T}^{\mu \nu}(0)$ is the energy-momentum tensor density operator of the hadron but this relation $<P| {\hat T}^{++}(0)|P>=2(P^+)^2$ is not correct if ${\hat T}^{\mu \nu}(0)$ is the gauge invariant color singlet energy-momentum tensor density operator of all the quarks plus antiquarks plus gluons inside the hadron. Hence we find that the momentum sum rule is violated in the operator product expansion (OPE) in QCD at high energy colliders.
\end{abstract}
\pacs{11.30.-j, 11.30.Cp, 11.15.-q, 12.38.-t }
\maketitle
\pagestyle{plain}

\pagenumbering{arabic}

\section{Introduction}

After the discovery of the renormalization \cite{rnvt} and the asymptotic freedom \cite{asd,asd1} in quantum chromodynamics (QCD) the prediction power of the perturbative QCD (pQCD) at short distance is tested at various collider and fixed target experiments. For the incoming and outgoing hadrons at the high energy colliders the factorization theorem in QCD \cite{factr} has played a major role along with the DGLAP evolution equation \cite{dgd} to make prediction of physical observable by using pQCD calculation.

One of the major problem remains to be solved in QCD is the non-perturbative QCD at log distance which will provide us information about hadron formation from quarks and gluons. Since the non-perturbative QCD is not solved yet the non-perturbative parton distribution function (PDF) inside the hadron and parton to hadron fragmentation function (FF) are extracted from various collider and fixed target experiments.

The momentum sum rule in QCD plays an important role at high energy colliders. In particular it is widely used in various parton distribution function (PDF) sets such as CTEQ \cite{cteq}, GRV \cite{grv}, MRST \cite{mrst} etc. for the phenomenological prediction of quark and gluon distribution functions inside the hadron for practical use. The momentum sum rule \cite{fcd,jnc} states that the momentum of all the quarks plus antiquarks plus gluons inside the hadron at high energy colliders is equal to the momentum of the hadron.

In order to see the practical importance of the momentum sum rule in QCD at high energy colliders consider the experimental measurement of the structure functions. Unlike quark structure function measurement it is not easy to measure gluon structure function at high energy experiments. Since the gluon structure function is not easily experimentally measured one can use the momentum sum rule to determine the gluon structure function from the experimentally measured quark structure function. This is one of the technique used in various PDF sets such as GRV \cite{grv}, MRST \cite{mrst} and CTEQ \cite{cteq} etc. to determine the quark and gluon distribution functions inside the hadron. Hence the momentum sum rule in QCD at high energy colliders has been very useful for practical purpose.

Since the momentum sum rule tells us that the momentum of all the quarks plus antiquarks plus gluons inside the hadron is equal to the momentum of the hadron, the momentum sum rule can be derived from the conservation of momentum in QCD from the first principle by using gauge invariant Noether's theorem. However, this is not straightforward like the momentum conservation statement in QED because of confinement in QCD, a phenomena which is absent in QED. The main point is that the momentum sum rule in QCD at high energy colliders equals the momentum of all the quarks plus antiquarks plus gluons inside the hadron to the momentum of the hadron. Since we do not know how the hadron is formed from quarks plus antiquarks plus gluons due to the lack of our knowledge about confinement in QCD which requires to solve non-perturbative QCD, we find that the proof of momentum sum rule in QCD from the first principle must include non-perturbative QCD at large distance where the hadron is formed.

Recently by using the gauge invariant Noether's theorem in QCD from the first principle we have proved that the momentum sum rule in QCD is violated at high energy colliders \cite{nkms} due to confinement in QCD which involves non-perturbative QCD at long distance. We have found non-vanishing boundary surface term at large distance in QCD where confinement happens because the potential energy at large distance $r$ in QCD is an increasing function of distance $r$ \cite{nkms}.

In the operator product expansion (OPE) in QCD the momentum sum rule is proved which does not need to use QCD at large distances \cite{jnc}. To prove the momentum sum rule in the operator product expansion (OPE) in QCD it is assumed that \cite{jnc}
\bea
<P| {\hat T}^{++}(0)|P>=2(P^+)^2
\label{hdnm}
\eea
where the ${\hat T}^{++}(0)$ is the $++$ component of the gauge invariant color singlet energy-momentum tensor density operator ${\hat T}^{\mu \nu}(0)$ of all the quarks plus antiquarks plus gluons inside the hadron $H$ and $|P>$ is the (physical) momentum state of the hadron $H$ with momentum $P^\mu$ normalized as
\bea
<P'| P>=2P^+~\delta(P'^+-P^+)~\delta^{(2)}({\bf P'_T}-{\bf P}_T).
\label{nrmcs}
\eea
In terms of $|{\tilde P}>$ which is normalized to unity we find from eq. (\ref{hdnm}) \cite{faj}
\bea
<P| {\hat T}^{++}(0)|P>=2P^+<{\tilde P}| \int d^3x ~{\hat T}^{++}(x)|{\tilde P}>= 2(P^+)^2.
\label{hdnmn}
\eea

However, in this paper, we show that this relation $<P| {\hat T}^{++}(0)|P>=2(P^+)^2$ in eq. (\ref{hdnm}) is correct if ${\hat T}^{\mu \nu}(0)$ is the energy-momentum tensor density operator of the hadron but this relation $<P| {\hat T}^{++}(0)|P>=2(P^+)^2$ in eq. (\ref{hdnm}) is not correct if ${\hat T}^{\mu \nu}(0)$ is the gauge invariant color singlet energy-momentum tensor density operator of all the quarks plus antiquarks plus gluons inside the hadron. 

This implies that the eq. (\ref{hdnm}) is an assumption in \cite{jnc} but is not a proof based on the first principle in QCD. Hence we find that the momentum sum rule is violated in the operator product expansion (OPE) in QCD at high energy colliders.

The paper is organized as follows. In section II we discuss the hadronic matrix element of hadronic operator. In section III we discuss the hadronic matrix element of partonic operator. In section IV we prove that the momentum sum rule is violated in the operator product expansion (OPE) in QCD at high energy colliders. Section V contains conclusions.

\section{ Hadronic Matrix Element Of Hadronic Operator }

Consider a particle of mass $M$ located at the space-time coordinate $X^\mu(t)$ with velocity $u^\mu=\frac{dX^\mu(t)}{dt}$, momentum $P^\mu$ and energy $P^0=E$. In classical mechanics, similar to the current density $j^\mu(x)$, we find that the energy-momentum tensor density $T^{\mu \nu}(x)$ of this particle is given by
\bea
T^{\mu \nu}(x)=\delta^{(3)}({\vec x}-{\vec X}(t))~\frac{P^\mu P^\nu}{P^0}.
\label{tmbd}
\eea
For free particle we find from eq. (\ref{tmbd})
\bea
\partial_\mu T^{\mu \nu}(x)=0
\label{tmbe}
\eea
which gives conserved energy-momentum
\bea
\int d^3x T^{0\mu}(x) = P^\mu.
\label{tmbf}
\eea

From eq. (\ref{tmbd}) we find the energy-momentum tensor 
\bea
\int d^3x T^{\mu \nu}(x)=\frac{1}{P^0}~P^\mu P^\nu.
\label{tmbg}
\eea
In quantum mechanics the momentum eigenvalue equation is given by
\bea
{\hat P}^\mu|P>=P^\mu |P>
\label{tmbl}
\eea
which gives
\bea
<P|{\hat P}^\mu|P>=P^\mu \int d^3x
\label{fa}
\eea
where $|P>$ is the momentum eigenstate with the normalization
\bea
<P'|P>=\delta^{(3)}({\vec P}'-{\vec P}).
\label{nrm}
\eea
Hence similar to eq. (\ref{hdnm}) if we use the momentum density operator ${\hat {\cal P}}^\mu(0)$ then we find from eq. (\ref{fa})
\bea
<P|{\hat {\cal P}}^\mu(0)|P>=P^\mu
\label{fb}
\eea
Extending the conserved momentum equation in classical mechanics from eq. (\ref{tmbf}) to quantum mechanics we find from eqs. (\ref{fb}) [similar to eq. (\ref{hdnmn})]
\bea
<P|{\hat {\cal P}}^\mu(0)|P>=< P| {\hat T}^{0\mu}(0)| P>=<{\tilde P}|\int d^3x {\hat T}^{0\mu}(x)|{\tilde P}>=P^\mu
\label{tmbfa}
\eea
which gives the conserved momentum $P^\mu$ where ${\hat T}^{\mu \nu}(x)$ is the energy-momentum tensor density operator and $|{\tilde P}>$ is normalized to unity.

In \cite{jnc} the normalization in the momentum eigenstate $|P>$ of the hadron in eq. (\ref{nrmcs}) differs from the normalization in the momentum eigenstate $|P>$ in eq. (\ref{nrm}) by a factor of $\sqrt{2P_0}$. Hence using the momentum eigenstate $|P>$ of hadron of \cite{jnc} from eq. (\ref{nrmcs}) we find from eq. (\ref{tmbfa}) that
\bea
<P|{\hat T}^{0 \mu}(0)|P>=2P_0<{\tilde P}|\int d^3x T^{0\mu}(x)|{\tilde P}> = 2P_0~P^\mu
\label{jca}
\eea
where $|{\tilde P}>$ is normalized to unity. Eq. (\ref{jca}) in the light-cone coordinate system gives
\bea
<P|{\hat T}^{++}(0)|P>=2P^+<{\tilde P}|\int d^3x {\hat T}^{++}(x)|{\tilde P}>=2~(P^+)^2
\label{tmbi}
\eea
which reproduces eq. (\ref{hdnm}) which agrees with \cite{jnc}.

\section{ Hadronic Matrix Element Of Partonic Operator }

Let us denote the momentum eigenstate of the gauge invariant color singlet momentum operator ${\hat p}^\mu_i$ of the parton $i=q,{\bar q},g$ by $|p_i>$ with the eigenvalue equation
\bea
{\hat p}^\mu_i|p_i>=p^\mu_i |p_i>
\label{tmbm}
\eea
where $p^\mu_i$ is the gauge invariant color singlet momentum of the parton $i$. Note that the momentum of the hadron in eq. (\ref{tmbl}) is denoted by capital $P^\mu$ whereas the momentum of the parton $i$ in eq. (\ref{tmbm}) is denoted by small $p^\mu_i$.

Hence we find from eqs. (\ref{tmbl}) and (\ref{tmbm}) that
\bea
{\hat p}^\mu_i |P>\neq P^\mu |P>
\label{tmbn}
\eea
where ${\hat p}^\mu_i$ is the gauge invariant color singlet momentum operator of the parton $i$, the $|P>$ is the (physical) momentum eigenstate of the hadron $H$ and $P^\mu$ is the momentum of the hadron $H$. The eq. (\ref{tmbn}) is true because momentum operator ${\hat p}^\mu_i$ of the parton $i$ is not physical because we have not directly experimentally observed quarks and gluons whereas $|P>$ and $P^\mu$ in eq. (\ref{tmbn}) are physical because $P^\mu$ and $|P>$ are the momentum and momentum eigenstate of the physical hadron.

The gauge invariant color singlet momentum operator ${\hat p}^\mu_i$ of a parton $i$ obtained from first principle by using gauge invariant Noether's theorem in QCD is given by
\bea
{\hat p}^\mu_i = \int d^3x {\hat T}^{0 \mu}_i(x)
\label{poi}
\eea
where ${\hat T}^{\mu \nu}_i(x)$ is the gauge invariant color singlet energy-momentum tensor density of the parton $i$ in QCD \cite{nkgi}.

It is important to mention here the important difference between the gauge invariant color singlet momentum operator ${\hat p}^\mu_i$ and momentum eigenstate $|p_i>$ of the parton $i$ in eq. (\ref{tmbm}) and the momentum operator ${\hat P}^\mu$ and momentum eigenstate $|P>$ of hadron in eq. (\ref{tmbl}). The momentum operator ${\hat P}^\mu$ and momentum eigenstate $|P>$ of hadron in eq. (\ref{tmbl}) are physical and hence the momentum is conserved, {\it i. e.},
\bea
\frac{d<P|{\hat {\cal P}}^\mu(0)|P>}{dt}=\frac{d<P| {\hat T}^{0\mu}(0)|P>}{dt}=\frac{d<{\tilde P}|\int d^3x {\hat T}^{0\mu}(x)|{\tilde P}>}{dt}=0.
\label{conv}
\eea
Hence it is important to remember that the momentum operator ${\hat P}^\mu$ of the hadron is a physically acceptable momentum operator because it is conserved, {\it i. e.}, it satisfies eq. (\ref{conv}).

However, the gauge invariant momentum operator ${\hat p}^\mu_i$ and momentum eigenstate $|p_i>$ of the parton $i$ in eq. (\ref{tmbm}) are not physical because we have not directly experimentally observed quarks and gluons. Hence it is not surprising that the momentum of the parton is not physical and hence is not conserved
{\it i. e.},
\bea
\frac{d<P|{\hat {\bar p}}^\mu_i(0)|P>}{dt}=\frac{d<P| {\hat T}^{0\mu}_i(0)|P>}{dt} =\frac{d<{\tilde P}|\int d^3x {\hat T}^{0\mu}_i(x)|{\tilde P}>}{dt} \neq 0
\label{conu}
\eea
where ${\hat {\bar p}}^\mu_i$ represents the momentum density operator of the parton $i$.

Note that even if the momentum operator ${\hat p}^\mu_i$ of the parton $i$ is color singlet and gauge invariant but the momentum $<P|{\hat {\bar p}}^\mu_i(0)|P>$ of the parton $i$ inside the hadron $H$ is not a physical observable because we have not directly experimentally observed quarks and gluons even if the momentum eigenstate $|P>$ of the hadron $H$ in eq. (\ref{conu}) is physical. Therefore it is important to remember that the momentum operator ${\hat p}^\mu_i$ of the parton is not a physically acceptable momentum operator because it is not conserved, {\it i. e.}, it satisfies eq. (\ref{conu}). As mentioned above, since we have not directly experimentally observed quarks and gluons, the eigenstate $|p_i>$ and the eigenvalue equation for the parton $i$ in eq. (\ref{tmbm}) does not correspond to any physical situation. We have just considered the eq. (\ref{tmbm}) as an illustration purpose to emphasize the validity of eq. (\ref{tmbn}).

From eqs. (\ref{tmbn}) and (\ref{poi}) we find
\bea
<P|{\hat T}^{0 \mu}_i(0)|P>=<{\tilde P}|\int d^3x {\hat T}^{0 \mu}_i(x)|{\tilde P}>\neq P^\mu.
\label{gind}
\eea
If we use the extra normalization factor $\sqrt{2P_0}$ in the (physical) momentum eigenstate $|P>$ of the hadron $H$ as used in eq. (\ref{nrmcs}) \cite{jnc} then we find from eq. (\ref{gind})
\bea
<P|{\hat T}^{0\mu}_i(0)|P>=2P_0<{\tilde P}|\int d^3x {\hat T}^{0 \mu}_i(x)|{\tilde P}>\neq 2P_0P^\mu
\label{gine}
\eea
where $|{\tilde P}>$ is normalized to unity. Eq. (\ref{gine}) in the light-cone coordinate system gives
\bea
<P|{\hat T}^{++}_i(0)|P>=2P^+<{\tilde P}|\int d^3x {\hat T}^{++}_i(x)|{\tilde P}>\neq 2(P^+)^2.
\label{ginf}
\eea

\section{ Momentum Sum Rule is Violated in the Operator Product Expansion in QCD at High Energy Colliders }

From the gauge invariant Noether's theorem \cite{nkgi} in QCD we find
\bea
<{\tilde P}|\sum_{i=q,{\bar q},g}\int d^3x \partial_\mu {\hat T}^{\mu \nu}_i(x)|{\tilde P}> =0
\label{gin}
\eea
where $|{\tilde P}>$ is the (physical) state of the hadron $H$ normalized to unity, the $\sum_{i=q,{\bar q},g}$ represents the sum of all the quarks plus antiquarks plus gluons inside the hadron $H$ and
\bea
&&{\hat T}^{\nu \mu}_g(x)= {\hat F}^{\nu \lambda c}(x) {\hat F}_\lambda^{~~\mu c}(x) +\frac{1}{4}g^{\nu \mu}  {\hat F}_{\lambda \delta}^c(x){\hat F}^{\lambda \delta c}(x)
\label{gigl}
\eea
is the gauge invariant energy-momentum tensor density operator of the gluon in QCD and
\bea
&&{\hat T}^{\nu \mu}_q(x)=\frac{i}{2} {\hat {\bar \psi}}_l(x)[\gamma^\nu  (\delta^{lk}{\overrightarrow \partial}^\mu -igT_{lk}^c{\hat Q}^{\mu c}(x)) -\gamma^\nu (\delta^{lk}{\overleftarrow \partial}^\mu +igT_{lk}^c{\hat Q}^{\mu c}(x))  ] {\hat \psi}_k(x)
\label{giqk}
\eea
is the gauge invariant energy-momentum tensor density operator of the quark in QCD. In eqs. (\ref{gigl}) and (\ref{giqk}) the ${\hat \psi}_i(x)$ is the quark field, ${\hat Q}_\mu^a(x)$ is the gluon field and
\bea
F_{\mu \delta}^c(x)=\partial_\mu {\hat Q}_\delta^c(x) -\partial_\delta {\hat Q}_\mu^c(x) +gf^{cdb} {\hat Q}_\delta^d(x) {\hat Q}_\mu^b(x).
\label{gfdb}
\eea
From eq. (\ref{gin}) we find
\bea
<{\tilde P}|\sum_{i=q,{\bar q},g}\int d^3x {\hat T}^{0 \mu}_i(x)|{\tilde P}>=-<{\tilde P} |\sum_{i=q,{\bar q},g} \int d^4x \partial_k {\hat T}^{k \mu}_i(x)]|{\tilde P}>
\label{ginb}
\eea
where in the $\int d^4x$ integration the $\int dt$ integration is an indefinite integration and $\int d^3x$ integration is definite integration.

Using eq. (\ref{poi}) in (\ref{ginb}) we find
\bea
<{\tilde P}|\sum_{i=q,{\bar q},g}\int d^3x {\hat T}^{0 \mu}_i(x)|{\tilde P}>=<P|\sum_{i=q,{\bar q},g}{\hat {\bar p}}^\mu_i(0) |P>=-<{\tilde P} |\sum_{i=q,{\bar q},g} \int d^4x \partial_k {\hat T}^{k \mu}_i(x)]|{\tilde P}>\nonumber \\
\label{ginc}
\eea
where ${\hat {\bar p}}^\mu_i$ represents the momentum density operator of the parton $i$ and $|P>$ is the momentum eigenstate of the hadron $H$ with momentum $P^\mu$ satisfying the normalization in eq. (\ref{nrm}) and $|{\tilde P}>$ is normalized to unity.

Note that if the boundary surface term does not vanish in QCD, {\it i. e.}, if
\bea
<{\tilde P} |\sum_{i=q,{\bar q},g} \int d^3x \partial_k {\hat T}^{k \mu}_i(x)]|{\tilde P}>\neq 0
\label{jhv}
\eea
then we find
\bea
<{\tilde P} |\sum_{i=q,{\bar q},g} \int d^4x \partial_k {\hat T}^{k \mu}_i(x)]|{\tilde P}>= {\rm time~dependent}
\label{khv}
\eea
which gives from eq. (\ref{ginc})
\bea
<{\tilde P}|\sum_{i=q,{\bar q},g}\int d^3x {\hat T}^{0 \mu}_i(x)|{\tilde P}>=<P|\sum_{i=q,{\bar q},g}{\hat {\bar p}}^\mu_i(0) |P>= {\rm time~dependent}.
\label{lhv}
\eea
Hence we find that if the boundary surface term does not vanish in QCD, {\it i. e.}, if the eq. (\ref{jhv}) is satisfied then we find from eq. (\ref{lhv}) that
\bea
<P|{\hat T}^{0\mu}(0)|P>=<P|\sum_{i=q,{\bar q},g}{\hat T}^{0\mu}_i(0)|P>=<{\tilde P}|\sum_{i=q,{\bar q},g}\int d^3x {\hat T}^{0 \mu}_i(x)|{\tilde P}>=<{\tilde P}|\sum_{i=q,{\bar q},g}{\hat {\bar p}}^\mu_i(0) |{\tilde P}>\neq P^\mu \nonumber \\
\label{mhv}
\eea
If we use the extra normalization factor $\sqrt{2P_0}$ in the (physical) momentum eigenstate $|P>$ of the hadron $H$ as used in eq. (\ref{nrmcs}) \cite{jnc} then we find from eq. (\ref{mhv})
\bea
&& <P|{\hat T}^{0\mu}(0)|P>=<P|\sum_{i=q,{\bar q},g}{\hat T}^{0\mu}_i(0)|P>=2P_0<{\tilde P}|\sum_{i=q,{\bar q},g}\int d^3x {\hat T}^{0 \mu}_i(x)|{\tilde P}>=< P|\sum_{i=q,{\bar q},g}{\hat {\bar p}}^\mu_i(0) |P>\nonumber \\
&&\neq 2P_0P^\mu 
\label{nhv}
\eea
which in the light-cone coordinate system gives
\bea
<P|{\hat T}^{++}(0)|P>=<P|\sum_{i=q,{\bar q},g}{\hat T}^{++}_i(0)|P>=2P^+<{\tilde P}|\sum_{i=q,{\bar q},g}\int d^3x {\hat T}^{++}_i(x)|{\tilde P}>\neq 2(P^+)^2.\nonumber \\
\label{ohv}
\eea
Note that we have found non-vanishing boundary surface term in QCD as given by eq. (\ref{jhv}) in \cite{nkms}. This implies that by using $<P|T^{++}(0)|P>\neq 2~(P^+)^2$ from eq. (\ref{ohv}) in eq. (3.28) of \cite{jnc} we find
\bea
\sum_i \int_0^1 d\xi ~\xi ~f_{i/H}(\xi) =\frac{1}{  2~(P^+)^2} <P|T^{++}(0)|P> \neq 1
\label{tmbx}
\eea
which proves that the momentum sum rule is violated in the operator product expansion (OPE) in QCD at high energy colliders where $f_{i/H}(\xi)$ is the parton distribution function (PDF) of the parton $i$ inside the hadron $H$ and $\xi$ is the longitudinal momentum fraction of the parton with respect to the hadron.

Note that the momentum sum rule in QCD involving parton distribution function inside nuclei also plays an important role to determine the initial condition for the quark-gluon plasma formation at RHIC and LHC heavy-ion colliders \cite{gg5,gg6,gg7,gg8}.

The relation $<P| {\hat T}^{++}(0)|P>=2(P^+)^2$ in eq. (\ref{hdnm}) in \cite{jnc} was based on the assumption that the total momentum of all the quarks plus antiquarks plus gluons inside the hadron is conserved. However, since the quark and gluon are not physical, {\it i. e.}, since we have not directly experimentally observed quarks and gluons, the momentum of a parton inside the hadron is not conserved as shown in eq. (\ref{conu}). Only when the total momentum of all the quarks plus antiquarks plus gluons inside the hadron is conserved then it becomes equal to the momentum of the hadron so that eqs. (\ref{conv}) and (\ref{hdnm}) are satisfied.

Hence we find that the momentum sum rule is violated in the operator product expansion (OPE) in QCD at high energy colliders. The correct momentum sum rule in QCD at high energy colliders is given by eq. (5) of \cite{nkms}.

\section{Conclusions}
To prove the momentum sum rule in the operator product expansion (OPE) in QCD at high energy colliders it is assumed that $<P| {\hat T}^{++}(0)|P>=2(P^+)^2$ where $|P>$ is the momentum eigenstate of the hadron $H$ with momentum $P^\mu$ and  ${\hat T}^{++}(0)$ is the $++$ component of the gauge invariant color singlet energy-momentum tensor density operator ${\hat T}^{\mu \nu}(0)$ of all the quarks plus antiquarks plus gluons inside the hadron $H$. However, in this paper, we have shown that this relation $<P| {\hat T}^{++}(0)|P>=2(P^+)^2$ is correct if ${\hat T}^{\mu \nu}(0)$ is the energy-momentum tensor density operator of the hadron but this relation $<P| {\hat T}^{++}(0)|P>=2(P^+)^2$ is not correct if ${\hat T}^{\mu \nu}(0)$ is the gauge invariant color singlet energy-momentum tensor density operator of all the quarks plus antiquarks plus gluons inside the hadron. Hence we have found that the momentum sum rule is violated in the operator product expansion (OPE) in QCD at high energy colliders.

\end{document}